\documentclass[fleqn]{article}
\usepackage{longtable}
\usepackage{graphicx}

\begin {document}
\begin{flushleft}
{\LARGE
{\bf Comment on ``Collision strength and effective collision strength for Br~XXVII" by  Goyal et al.  [Can. J. Phys.   95 (2017) 1127]}
}\\

\vspace{1.5 cm}

{\bf {Kanti  M  ~Aggarwal}}\\ 

\vspace*{1.0cm}

Astrophysics Research Centre, School of Mathematics and Physics, Queen's University Belfast, \\Belfast BT7 1NN, Northern Ireland, UK\\ 
\vspace*{0.5 cm} 

e-mail: K.Aggarwal@qub.ac.uk \\

\vspace*{0.20cm}

Received: 2 November 2017; Accepted: 23 May 2018

\vspace*{1.0 cm}

\vspace*{1.0 cm}
\vspace*{1.0 cm}

\hrule

\vspace{0.5 cm}

\end{flushleft}

\clearpage


\begin{abstract}

In a recent paper, Goyal et al. [Can. J. Phys.  {\bf 95} (2017) 1127] have reported results for  collision strengths  ($\Omega$) and effective collision strengths ($\Upsilon$) for transitions among the lowest 52 levels of F-like Br~XXVII. For their calculations, they have adopted the Dirac atomic $R$-matrix code (DARC) and the flexible atomic code (FAC). In this comment we demonstrate that their results for both parameters are erratic, inaccurate, and unreliable.   

\end{abstract}

\clearpage

\section{Introduction}

In a recent paper, Goyal et al. \cite{mm1} have reported results for  collision strengths  ($\Omega$) and effective collision strengths ($\Upsilon$) for transitions among the lowest 52 levels of F-like Br~XXVII. These levels belong to the 2s$^2$2p$^5$, 2s2p$^6$, and 2s$^2$2p$^4$3$\ell$ configurations. For their calculations, they have adopted the Dirac atomic $R$-matrix code (DARC) and the flexible atomic code (FAC), which are both available on the websites \\{\tt http://amdpp.phys.strath.ac.uk/UK\_APAP/codes.html} and {\tt https://www-amdis.iaea.org/FAC/}, respectively. They have concluded a good agreement between the two sets of results, for a majority of transitions. However, we do not find this to be true and in addition,  demonstrate that their results for both parameters are erratic, inaccurate, and unreliable.   

For the calculations of $\Omega$, and subsequently $\Upsilon$, Goyal et al. \cite{mm1} have included 27 configurations in the construction of their wavefunctions, which  belong to  2s$^2$2p$^5$, 2s2p$^6$, 2p$^6$3$\ell$, 2s$^2$2p$^4$3$\ell$, 2s2p$^5$3$\ell$, 2s$^2$2p$^4$4$\ell$, 2s2p$^5$4$\ell$, 2s$^2$2p$^4$5$\ell$, and 2s2p$^5$5$\ell$, with $\ell \le$ f, and generate 431 levels in total.  For these calculations, the GRASP0 version of the general-purpose relativistic atomic structure package of P.H.~Norrington and I.P.~Grant has been adopted, which is available at the same website as DARC. However, for the scattering calculations they have considered only the lowest 52 levels, which belong to the 2s$^2$2p$^5$, 2s2p$^6$, and 2s$^2$2p$^4$3$\ell$  configurations. In fact, these configurations generate 60 levels in total, but they have preferred to ignore the remaining 8, perhaps for the computational reason. Similarly, it is not clear exactly how many levels/configurations they have adopted in their other calculations with FAC. However for comparison purpose,  we include 113 levels of the 2s$^2$2p$^5$, 2s2p$^6$, 2s$^2$2p$^4$3$\ell$,  2s2p$^5$3$\ell$, and 2p$^6$3$\ell$ configurations, i.e. 11 in total, in our calculations with FAC  to determine various atomic parameters.  In addition, we have also performed similar calculations with the GRASP code, adopting the same version as by them.

\section {Collision strengths}

Our level energies obtained with the GRASP and FAC codes are comparable with those reported by Goyal et al. \cite{mm1}, and hence are not listed here.  Anyway, our focus is on the calculations for $\Omega$ and $\Upsilon$. In Fig.~1, we show a comparison between the $\Omega$ obtained in our FAC calculations with those listed by Goyal et al. \cite{mm1} with DARC, for three {\em allowed} transitions, namely 4--14 (2s$^2$2p$^4$3s~$^4$P$_{5/2}$ -- 2s$^2$2p$^4$3p~$^4$D$^o_{7/2}$: triangles), 6--18 (2s$^2$2p$^4$3s~$^2$S$_{1/2}$ -- 2s$^2$2p$^4$3p~$^4$D$^o_{3/2}$: diamonds), and 8--21 (2s$^2$2p$^4$3s~$^4$P$_{3/2}$ -- 2s$^2$2p$^4$3p~$^4$P$^o_{5/2}$: stars) -- see Table~1 of \cite{mm1} for definition of all 52 levels.  These transitions have been chosen because of their comparatively large magnitude, and allowed transitions are much easier to compare with because their  $\Omega$  are strongly dependent on their oscillator strengths (f-values) and energy differences, i.e. $\Delta$E$_{ij}$.  

Not only Goyal et al. \cite{mm1} have performed their calculations in a very limited energy range, but their $\Omega$ results also show an incorrect behaviour as is clearly seen in Fig.~1. Their $\Omega$ are nearly constant with increasing energy, for all these three transitions (and many more), whereas those with FAC show an increasing trend, as expected. Their $\Omega$ values are {\em underestimated} by (almost) a factor of two. This is in spite of including the contribution of higher neglected partial waves (i.e. top-up) through a Coulomb-Bethe approximation, as claimed by them. It is clear from this figure that either they have not included the top-up for allowed transitions, or their procedure is wrong. In any case, it does not support their `conclusion' that there is a good agreement between the FAC and DARC results for $\Omega$, for most of the transitions. We discuss this further. 

\begin{figure*}
\includegraphics[angle=-90,width=0.9\textwidth]{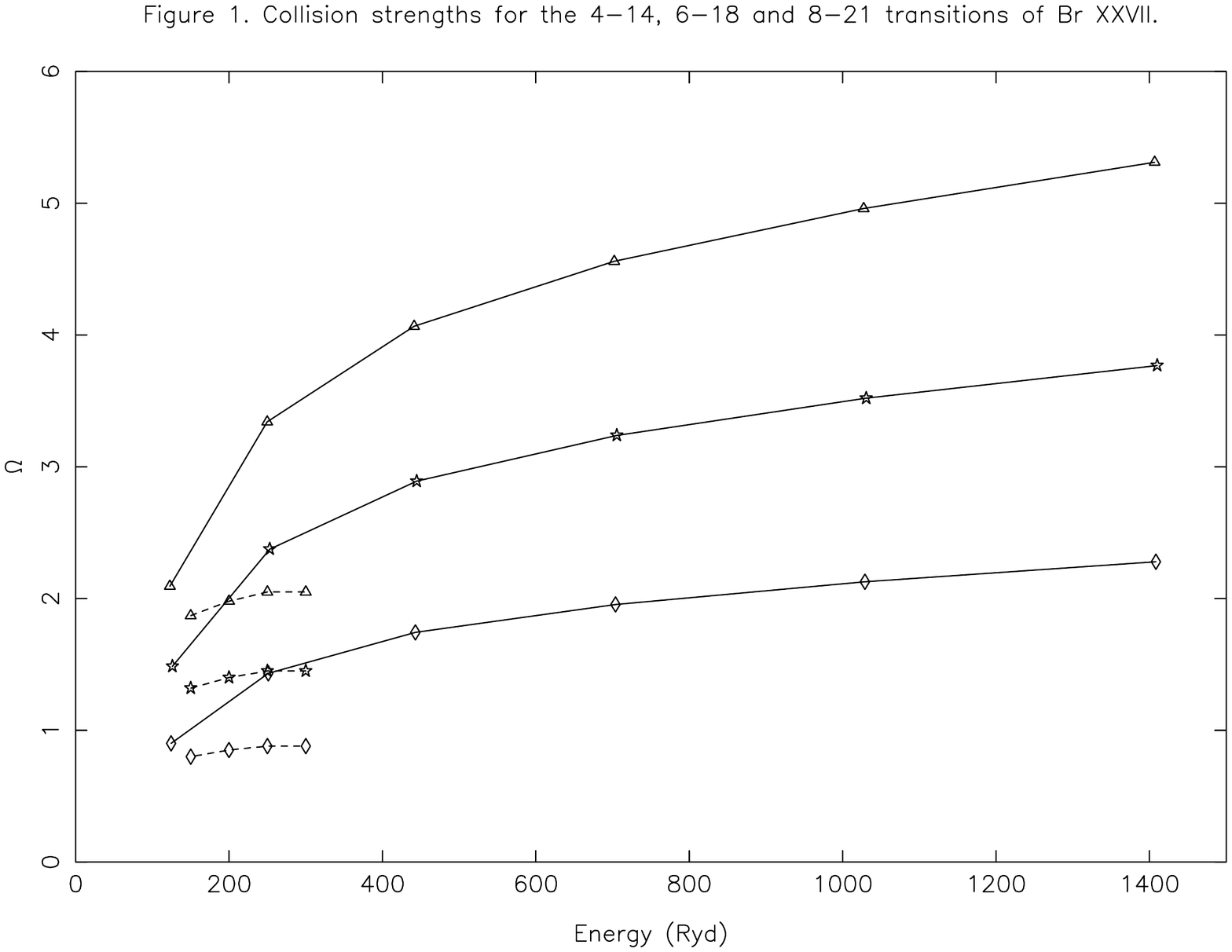}
 \vspace{-1.5cm}
 \caption{Comparison of DARC and FAC values of $\Omega$ for  the 4--14 (triangles), 6--18 (diamonds), and 8--21 (stars) transitions of Br~XXVII. Continuous curves: present results with FAC, broken curves: earlier results of Goyal et al. \cite{mm1} with DARC.}
 \end{figure*}
 
 \begin{figure*}
\includegraphics[angle=-90,width=0.9\textwidth]{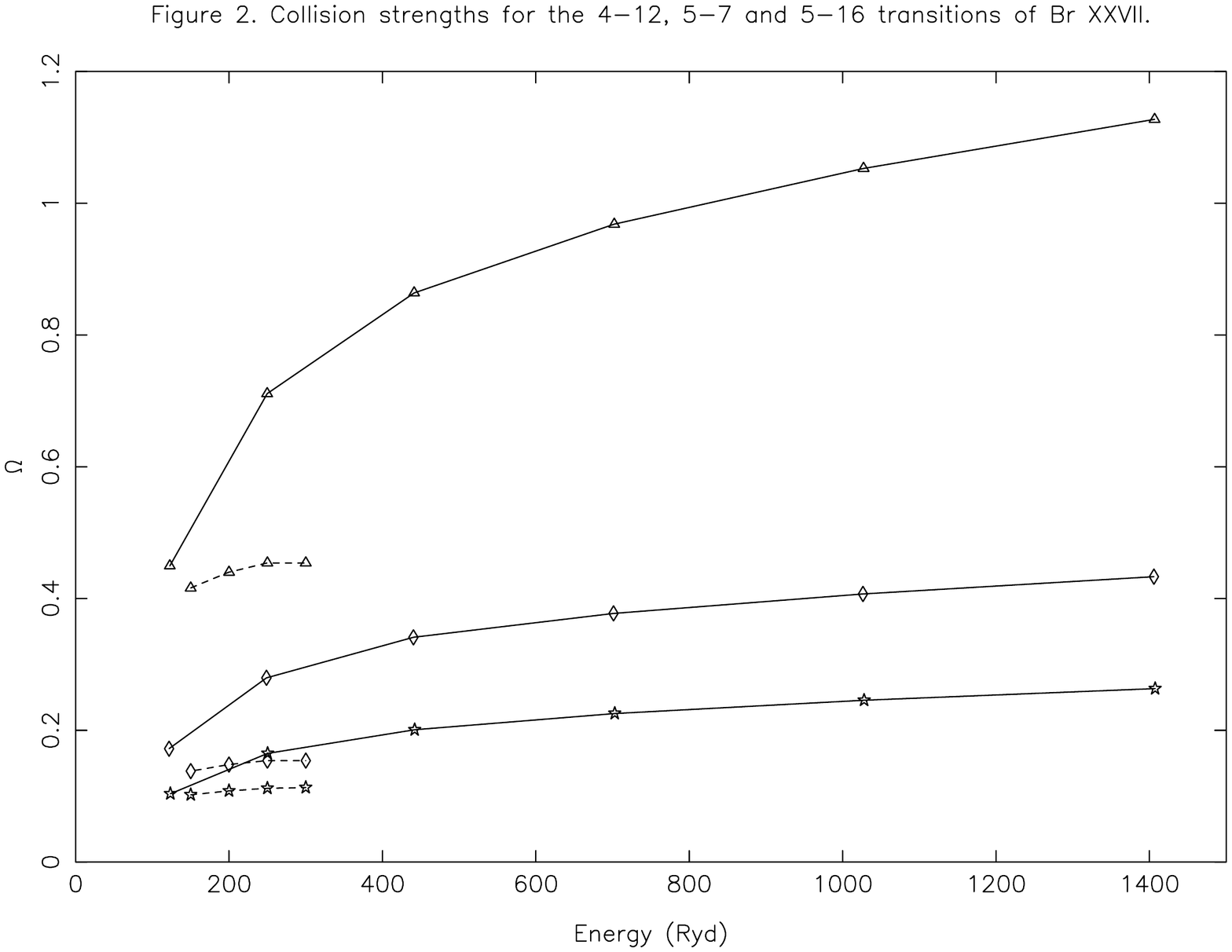}
 \vspace{-1.5cm}
 \caption{Comparison of DARC and FAC values of $\Omega$ for  the 4--12 (triangles), 5--7 (diamonds), and 5--16 (stars) transitions of Br~XXVII. Continuous curves: present results with FAC, broken curves: earlier results of Goyal et al. \cite{mm1} with DARC.}
 \end{figure*}

In Fig.~2, we show a similar comparison for three other allowed transitions, namely 4--12 (2s$^2$2p$^4$3s~$^4$P$_{5/2}$ -- 2s$^2$2p$^4$3p~$^2$D$^o_{5/2}$: triangles), 5--7 (2s$^2$2p$^4$3s~$^2$P$_{3/2}$ -- 2s$^2$2p$^4$3p~$^4$P$^o_{3/2}$: diamonds), and 5--16 (2s$^2$2p$^4$3s~$^2$P$_{3/2}$ -- 2s$^2$2p$^4$3p~$^4$D$^o_{1/2}$: stars). These are {\em inter-combination} transitions, but allowed in the $jj$ coupling scheme, adopted in both the DARC and FAC calculations. As for the transitions in Fig.~1, these three also have neither the right behaviour nor the correct magnitude. It may be worth noting here that the version of DARC available on the website, and adopted by Goyal et al. \cite{mm1}, does not have a provision of the Coulomb-Bethe approximation for the allowed transitions, and how they have performed their calculations is unclear.

Although the $\Omega$ behaviour of the DARC calculations by Goyal et al. \cite{mm1} is clearly incorrect for the allowed transitions, as seen in Figs.~1 and 2, one may have a question about the magnitude, because as already stated, it depends on the f-value and $\Delta$E$_{ij}$. Since the energies obtained in both the GRASP and FAC calculations are similar and comparable for almost all levels, we show in Table~1 that the f-values are also comparable, for all the six transitions shown in Figs. 1 and 2, and many more. Since Goyal et al. have not listed their f-values, we have performed another calculation with GRASP by including the same 431 levels as by them. Therefore, listed in Table~1 are three sets of f-values, namely (i) GRASP1, which includes 113 levels as adopted in the subsequent calculations of $\Omega$ and $\Upsilon$, (ii) GRASP2, which includes a larger CI (configuration interaction) among 431 levels, as adopted by Goyal et al. in the construction of their wavefunctions, and finally (iii) FAC, with the same 113 levels as in GRASP1. For all transitions, the f-values are (practically) the same under different models and codes, and therefore the corresponding $\Omega$ values are expected to be similar and comparable. Unfortunately, this is not the case and the $\Omega$ results listed by Goyal et al. \cite{mm1} are {\em incorrect}, at least for the allowed transitions. 

\begin{table}
\caption{Comparison of oscillator strengths (f-values) for some transitions of Br~XXVII.} 
\begin{tabular}{rrcrrrr}  \hline
 I  &  J & Transition  & GRASP1 & GRASP2 & FAC  \\
 \hline
    4  &   12  &  2s$^2$2p$^4$3s~$^4$P$_{5/2}$ -- 2s$^2$2p$^4$3p~$^2$D$^o_{5/2}$ &  0.0252  &  0.0249  &  0.0250    \\
    4  &   14  &  2s$^2$2p$^4$3s~$^4$P$_{5/2}$ -- 2s$^2$2p$^4$3p~$^4$D$^o_{7/2}$ &  0.1184  &  0.1180  &  0.1192    \\
    5  &    7  &  2s$^2$2p$^4$3s~$^2$P$_{3/2}$ -- 2s$^2$2p$^4$3p~$^4$P$^o_{3/2}$ &  0.0092  &  0.0092  &  0.0092    \\
    5  &   16  &  2s$^2$2p$^4$3s~$^2$P$_{3/2}$ -- 2s$^2$2p$^4$3p~$^4$D$^o_{1/2}$ &  0.0091  &  0.0092  &  0.0094    \\
    6  &   18  &  2s$^2$2p$^4$3s~$^2$S$_{1/2}$ -- 2s$^2$2p$^4$3p~$^4$D$^o_{3/2}$ &  0.1567  &  0.1577  &  0.1567    \\
    8  &   21  &  2s$^2$2p$^4$3s~$^4$P$_{3/2}$ -- 2s$^2$2p$^4$3p~$^4$P$^o_{5/2}$ &  0.1224  &  0.1219  &  0.1229    \\
\hline	
\end{tabular}

\begin{flushleft}
{\small
GRASP1: calculations with GRASP for 113 levels \\
GRASP2:  calculations with GRASP for 431 levels \\
FAC: calculations with FAC for 113 levels  \\
}
\end{flushleft}
\end{table}

\section {Effective collision strengths}
 
\begin{figure*}
\includegraphics[angle=-90,width=0.9\textwidth]{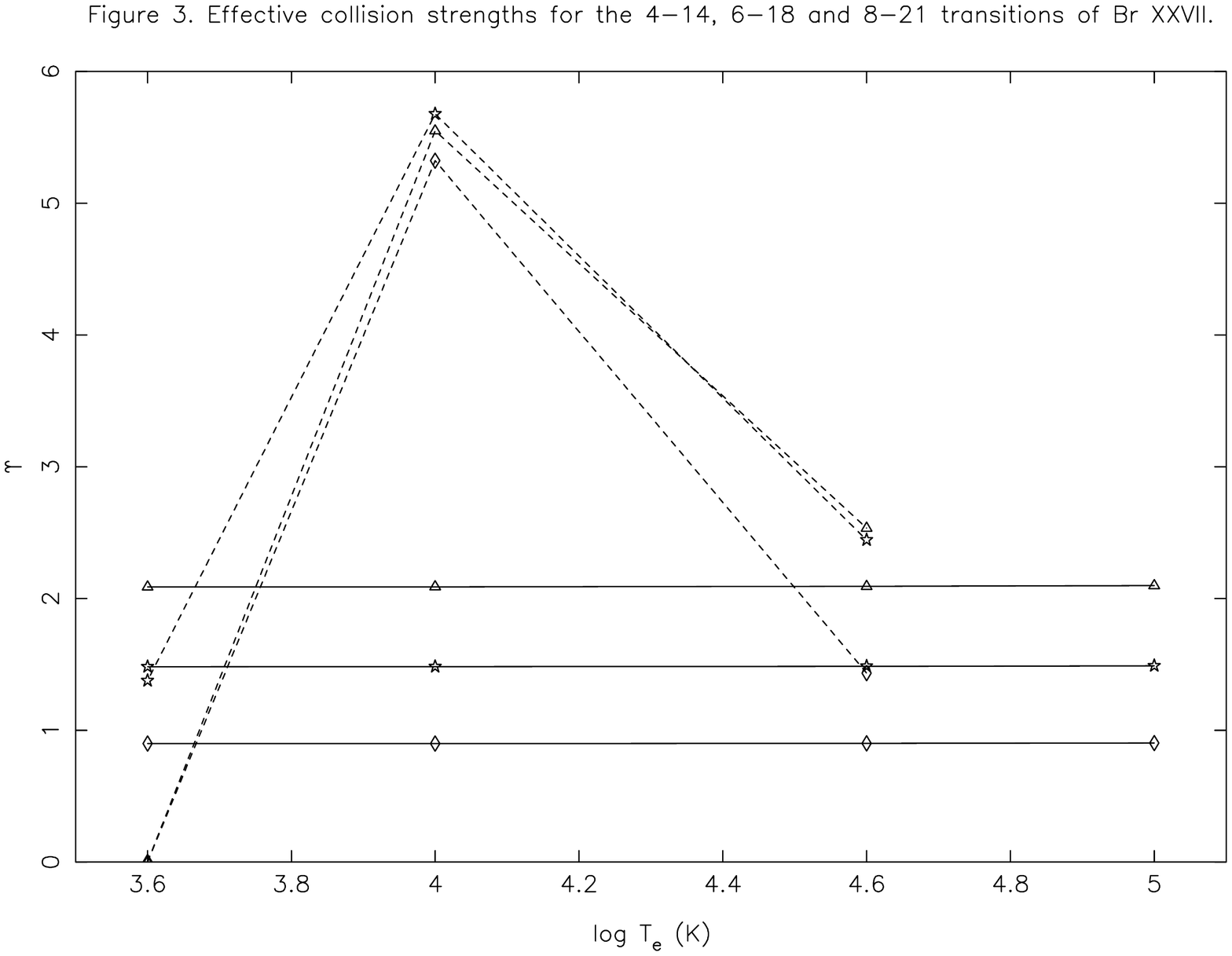}
 \vspace{-1.5cm}
 \caption{Comparison of DARC and FAC  values of $\Upsilon$ for the 4--14 (triangles), 6--18 (diamonds), and 8--21 (stars) transitions of Br~XXVII. Continuous curves: present results with FAC, broken curves: earlier results of Goyal et al. \cite{mm1} with DARC.}
 \end{figure*}

 \begin{figure*}
\includegraphics[angle=-90,width=0.9\textwidth]{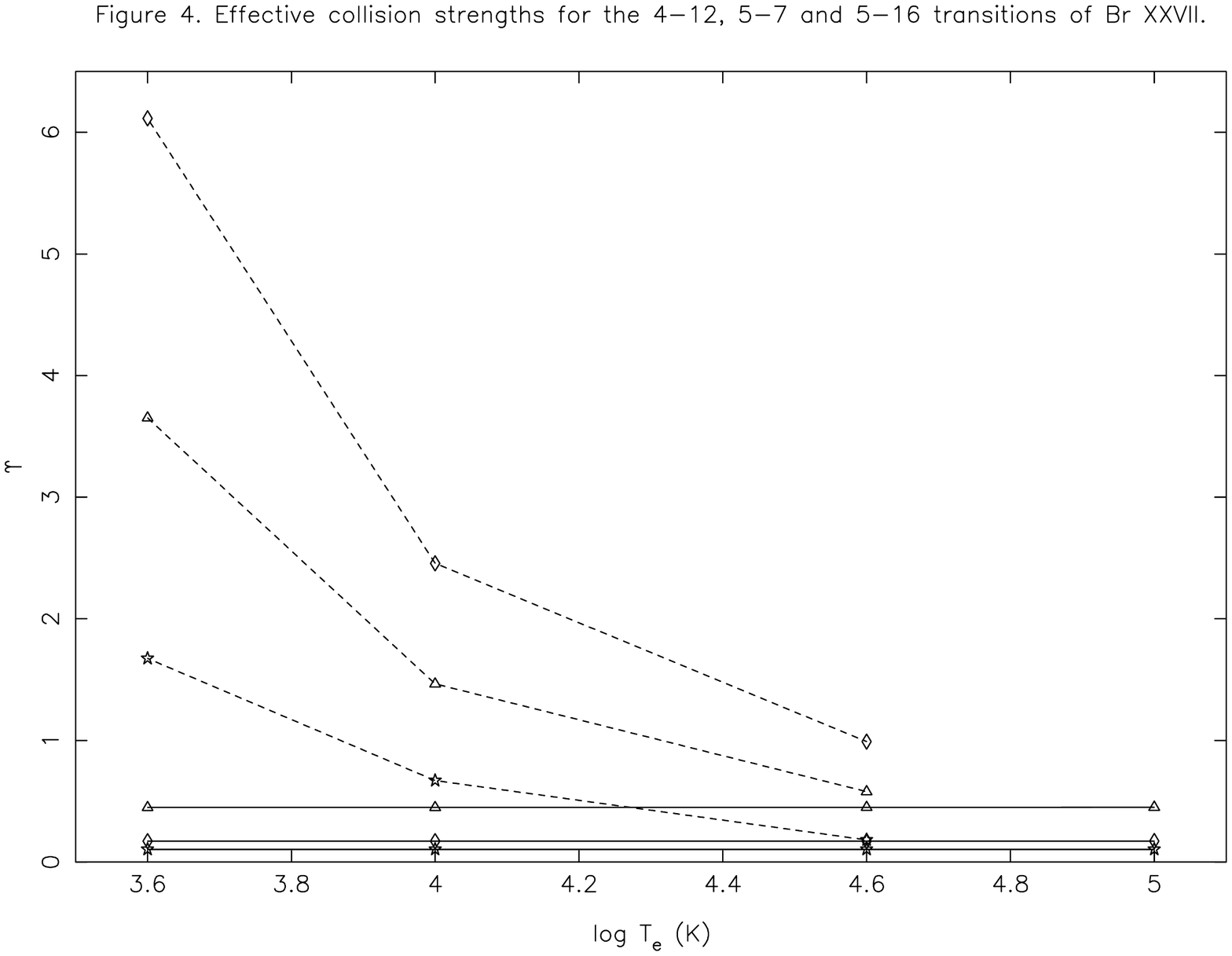}
 \vspace{-1.5cm}
 \caption{Comparison of DARC and FAC  values of $\Upsilon$ for the 4--12 (triangles), 5--7 (diamonds), and 5--16 (stars) transitions of Br~XXVII. Continuous curves: present results with FAC, broken curves: earlier results of Goyal et al. \cite{mm1} with DARC.}
 \end{figure*}

 For the calculations of $\Upsilon$, Goyal et al. \cite{mm1} have determined $\Omega$ with the DARC code up to an energy of 375~Ryd, have included partial waves with angular momentum $J \le$ 41, and have resolved resonances. They have reported results for all possible 1326 transitions among the lowest 52 levels of Br~XXVII, and at three temperatures of 4000, 10~000, and 40~000~K. Unfortunately, their reported results for $\Upsilon$ are in greater error than for $\Omega$. This is very clear from a simple look at their Table~3, because for three transitions (1--22, 1--42, and 1--49) their $\Upsilon$ are 0.0 at T$_e$ = 4000~K, and this is irrespective of the magnitude of their $\Omega$ or the $\Upsilon$ results at other two temperatures. In their supplementary table, there are many more such transitions, and examples include: 2--22/42/49, 3--19/22/29, and 4--22/42/49. For some transitions, such as 28--49, 32--49, and 33--49, their $\Upsilon$ are 0.0 at more than one T$_e$, and for a few these are simply ****, see for example: 12--33, 13--34, and 14--32.  We discuss this further below.

Since forbidden transitions normally exhibit a range of resonances, as shown in Figs.~4--10 of \cite{mm1}, it is much easier to compare the $\Upsilon$ results, between any two calculations, for the allowed ones, because resonances for these are neither numerous nor very high in magnitude. Therefore effectively, the contribution of resonances in the determination of $\Upsilon$ is not very appreciable. For this reason, in Fig.~3 we compare the $\Upsilon$ results for the same three transitions as in Fig.~1, i.e. 4--14, 6--18, and 8--21. Clearly, the $\Upsilon$ of Goyal et al. \cite{mm1} are erratic in behaviour and incorrect in magnitude. The temperature of 40~000~K, the highest considered by them, corresponds to only 0.25~Ryd, and in this small energy/temperature range the $\Upsilon$ values do not change drastically. As seen in the corresponding results with FAC, these are almost constant, as expected.

A closer look at Figs.~5 and 8 of \cite{mm1} for the 1--3 (2s$^2$2p$^5$~$^2$P$^o_{3/2}$ -- 2s2p$^6$~$^2$S$_{1/2}$) and 2--3 (2s$^2$2p$^5$~$^2$P$^o_{1/2}$ -- 2s2p$^6$~$^2$S$_{1/2}$) transitions, both allowed,  reveals that their $\Omega$ vary smoothly in the lowest $\sim$20~Ryd energy range,  without any resonances. This means that at T$_e$ below 40~000~K, the $\Upsilon$ should also vary smoothly, if not (nearly) constant. Our calculations with FAC confirm this as the $\Upsilon$ are about 0.15 and 0.08, for the respective two transitions. The corresponding results of Goyal et al. \cite{mm1} are 1.054, 0.425, and 0.193 for 1--3, and 0.577, 0.233, and 0.105 for 2--3, at the respective temperatures of 4000, 10~000, and 40~000~K. Clearly, these results are simply wrong. This may be further confirmed by our calculations with DARC for another F-like ion \cite{kr28}, namely Kr~XXVIII which is close to Br~XXVII. As seen in Table~3 of \cite{kr28}, the $\Upsilon$ values for these two transitions are nearly constant at temperatures below 10$^6$~K. 

In Fig.~4, we show similar comparison of $\Upsilon$ results for other three transitions (4--12, 5--7, and 5--16), which are the same as seen in Fig.~2. For these three transitions (and many more) their $\Upsilon$ values do not appear to be correct, particularly at lower temperatures. Our $\Upsilon$ results with FAC show a slight increase with increasing T$_e$, which follows from the increasing $\Omega$ with energy for these, whereas those of \cite{mm1} decrease, apart from an anomalous behaviour at low T$_e$. The main reason for this erratic, anomalous, and incorrect behaviour of $\Upsilon$ results of Goyal et al. \cite{mm1} is, in our opinion, their choice of energy mesh ($\delta$E) for the resolution of resonances and the determination of $\Upsilon$. Their $\delta$E is 0.065~Ryd which amounts to 0.88~eV, or equivalently $\sim$10~260~K. To calculate $\Upsilon$ at low temperatures, such as 4000~K, it needs to be in the range of (not more than) 50 to 100~K. In fact to calculate $\Upsilon$ up to 40~000~K (0.25~Ryd), results for $\Omega$ should be sufficient for only 1~Ryd of energy. Hence, there was no apparent requirement to calculate $\Omega$ up to 375~Ryd, i.e. $\sim$245~Ryd {\em above} the highest threshold.

\section{Conclusions}

In this short paper, we have demonstrated that the $\Omega$ and $\Upsilon$ results reported by Goyal et al. \cite{mm1} for transitions in F-like Br~XXVII are erratic and  anomalous in behaviour, and incorrect in magnitude, and therefore completely unreliable. Their $\Omega$ results are underestimated for the allowed transitions, because of either non inclusion of the contribution of higher neglected partial waves, or the wrong procedure applied by them. The erratic behaviour of their $\Upsilon$ results is because of the large energy mesh adopted by them to make calculations at very low temperatures. In addition, their assertion of a good agreement between the DARC and FAC calculations of $\Omega$ for a majority of transitions is not true. They have arrived at this conclusion by making comparisons for less than 4\% of the transitions, and that too at a single energy. Such limited comparisons, as also noted recently for transitions in Mg~V \cite{mgv}, often lead to wrong conclusions and incorrect results. Furthermore, their results for another F-like ion, namely W~LXVI, suffer from similar deficiencies,  and contain large errors  in various atomic parameters, including $\Omega$, as has been recently explained and demonstrated by us \cite{w66a},\cite{w66b}. Most of the time large errors (read discrepancies) in reported atomic data are because of the insufficient comparisons made by the producers of data. Many such examples, with possible reasons and suggestions for improvement, have been listed in a recent paper by us \cite{atom}.

In view of our above observations and assessment, it is recommended that fresh calculations should be performed for the collisional data for Br~XXVII. Additionally, the likely applications for this data are for the modelling of fusion plasmas for which much higher temperatures, of the orders of 10$^6$ to 10$^8$~K, prevail. Similarly, a larger range of transitions, covering at least 113 levels of the 2s$^2$2p$^5$, 2s2p$^6$, 2s$^2$2p$^4$3$\ell$,  2s2p$^5$3$\ell$, and 2p$^6$3$\ell$ configurations, should be considered, as was the case for Kr~XXVIII~\cite{kr28}.



\newpage
\begin{thebibliography}{999}

\bibitem{mm1} A. Goyal, R. Sharma, I. Khatri, A.K. Singh, S.S. Singh, and M. Mohan. Can. J. Phys.  {\bf 95}, 1127 (2017).
\bibitem{kr28} K.M. Aggarwal, F.P.  Keenan, and K.D. Lawson. At. Data Nucl. Data Tables {\bf 97}, 225 (2011).
\bibitem{mgv} K.M. Aggarwal and F.P.  Keenan. Can. J. Phys. {\bf 95}, 9 (2017).
\bibitem{w66a} K.M. Aggarwal.  Chin. Phys. B {\bf 25},  043201 (2016).
\bibitem{w66b} K.M. Aggarwal. Atoms {\bf 4}, 4030024 (2016).
\bibitem{atom} K.M.  Aggarwal. Atoms {\bf 5}, 5040037 (2017). 

\end{thebibliography}
\end{document}